\documentclass{jkas}
\def\year{2019} 
\def\month{August} 
\def\volume{00} 
\def\issue{0} 
\def\beginpage{1} 
\def\endpage{18} 
\def\received{April 30, 2019} 
\def\accepted{July 23, 2019} 
\date{Received \received; accepted \accepted}

\title{Economic Power, Population, and the Size of Astronomical Community}

\author[1]{Sang-Hyeon~Ahn}

\affil[1]{Center for Theoretical Astronomy, Korea Astronomy and Space Science Institute, Daedeokdaero 776, Yuseong-gu, Daejeon 34055, Republic of Korea; \email{sha@kasi.re.kr}}

\def\corrauthor{Sang-Hyeon Ahn}
\def\runningauthor{S. -H. Ahn}
\def\runningtitle{Eeconomic Power and the size of astronomical community}
\def\keywords{General: sociology of astronomy, methods: data analysis, statistical}

\def\abstracttext{The number of astronomers for a country registered to the International Astronomical Union (IAU) is known to have a correlation with the gross domestic product (GDP). However, the robustness of this relationship can be doubted, because the fraction of astronomers joining the IAU differs from country to country. Here we revisit this correlation by using the recent data updated as of 2017, and then we find a similar correlation by using the total enumeration of astronomers and astrophysicists with PhD degrees and working in each country, instead of adopting the number of IAU members. We confirm the existence of the correlation. We also confirm the existence of two subgroup in the correlation. One group consists of European advanced countries having long history of modern astronomy, while the other group consists of countries having experienced recent rapid economic development. In order to find causation in the correlation, we obtain the long-term variations of the number of astronomers, population, and the GDP for a number of countries to find that the number of astronomers per citizen for recently developing countries has increased more rapidly as GDP per capita increased, than that for fully developed countries. We collect a demographic data of the Korean astronomical community to find that it has experienced a recent rapid growth. From these findings we estimate the proper size of the Korean astronomical community by considering the society’s economic power and population. The current number of PhD astronomers working in Korea is approximately 310, but it should be 550 that is large enough to be comparable and competitive to the sizes of Spainish, Canadian, and Japanese astronomical communities, which will be able to be reached by the year of 2030 if the current increasing trend continues. The proper number of PhD astronomers in Korea should be 780 in order to be comparable to the German, French, and Italian communities. We discuss on the way how to overcome the vulnerability of the Korean astronomical community, based on the statistics of national R\&D expenditure structure comparing with that of other major advanced countries.}

\begin{document}

\jkashead

\section{Introduction \label{sec:intro}}

Scientific curiosity of human beings has long been laid on wonders of the Universe, which have stimulated the intelligent to develop modern astronomy as being one of the fascinating subjects. Being a pure science, the size of astronomical community of a country can be a measure of scientific development of the country. It has been one of highly interesting topics to governments, as well as scientific communities, to know a nation's position among the international science communities. Thus, analyzing the astronomical communities can give us some lesson on this issue.

There are a large number of previous researches on the relation between economic capability and scientific research output, in either quantitative or qualitative aspects. 
For example, \citet{may97} analyzed the scientific research outputs among several countries, based on the Science Citation Index established by the Institute for Scientific Information. He found the existence of the large differences in performance among nations, which was ascribed to differences in the nature of the institutional settings between Germany/France and the UK/USA/Scandinavian countries. He stressed that the latter performed better because of their heritage of nonhierarchical nature and continuous stimulations injected by young generation. Similar approaches on this topic were also made by \citet{king04}, \citet{moed05}, and \citet{hohmann17}, who performed bibliometric analyses for publication/citation rates. 

The NASA Astrophysics Data System (ADS) Abstract Service was first demonstrated in 1992 and was put on-line service for general use in 1993. This system is widely used in the international astronomical communities, and so provides the best data for the bibliometric researches. \citet{kurtz05} analyzed the NASA ADS query data in detail. They found a few  things. First of all, when we consider the intensities rather than total volumes, the largest per capita user of the ADS is not the USA or the largest community, but the France and the Netherlands. Second of all, they found that while the difference in per capita income between the richest and poorest countries is a factor of ten to fifteen, the difference in per capita ADS use reaches roughly a factor of three hundred. Last of all, eastern European countries perform efficiently astronomical resarches by using ADS system, which means that GDP is not a proper measure of the wealth of nations when they are undergoing substantial economic and political changes.
\citet{henneken19} introduced the reading activity or query frequency to the ADS in attempts to quantify the research efficiency, as well as the traditional bibliometric indicators, like publication and citation. 
  
We agree on an idea that the astronomy can be regarded as a proxy for all basic sciences beacause there is no applied astronomy, as was pointed out by \citet{kurtz05}. To prove this idea is beyond the scope of this paper. However, admitting the assumption, we will only focus on the astronomical community only rather than consider all the entire scientific communities. Additionally, we will not consider the bibliometric data, which usually indicates the output or performance of research activities. Instead, in this paper, we will just concentrate on the traditional indicators such as the number of astronomers and the GDP, in either volumes or intensities. Nevertheless, we note that this work is meaningful, because we will use the most recent data and also the time series data that can trace back to 30 years ago.

\section{Data}
One indicator for the characteristics of astronomical communities is the number of astronomers registered to the International Astronomical Union (hereafter IAU in abbreviation). It is known that the number has a correlation with the gross domestic product (hereafter abbreviated as GDP), which is often regarded as an indicator of economic power \citep{hearnshaw01, hearnshaw06, ribeiro13}. However, we can easily agree on an idea that the percentage of astronomers joining the IAU varies from country to country. Thus, we need to consider the full volume of the astronomers in this paper. 

We adopt the GDP per capita, the GDP purchasing power parity (PPP in abbreviation) in current USD, and populations as of the year 2017 provided in the World Bank database\footnote{The World Bank national account data, and OECD National Accounts data files: \url{https://data.worldbank.org/indicator/ny.gdp.mktp.cd}}. The number of IAU members for each country is obtained from the statistics provided on the IAU website\footnote{International Astronomical Union. Geographical and Gender Distribution of Individual Members, \url{https://www.iau.org/administration/membership/individual/distribution/}.}. Table~\ref{tab:jkastab1} shows the number of IAU members, the GDP per capita, and population for each country.

\begin{table*}
\caption{{\bf GDP per capita, populations, and the number of astronomers joining into IAU.} Exceptionally, the GDP per capita of the Democratic People’s Republic of Korea (or North Korea) is not provided by IMF or World Bank. Thus we choose the figure of 1,360 USD estimated by Statistics Korea of the government of the Republic of Korea. The GDP per capita for Egypt is also not provided for the year 2017, and so we get the value in the year 2016.\label{tab:jkastab1}}
\centering
\begin{tabular}{clrrr|clrrr}
\toprule
No. & Country & GDP/cap. & Population & IAU & No. & Country & GDP/cap. & Population & IAU \\
 & & USD & million & members &  & & USD & million & members \\
\midrule
1 & USA & 59,495 & 326.6 & 2,824 & 31 & China & 8,583 & 1,379.3 & 663 \\
2 & Japan & 38,550 & 126.5 & 728 & 32 & Taiwan & 24,227 & 23.5 & 75 \\
3 & France & 39,673 & 67.1 & 856 & 33 & Argentina & 14,061 & 44.3 & 148 \\
4 & Germany & 44,184 & 80.6 & 654 & 34 & Australia & 56,135 & 23.2 & 328 \\
5 & India & 1,852 & 1,281.9 & 281 & 35 & Austria & 46,436 & 8.8 & 64 \\
6 & Indonesia & 3,859 & 260.6 & 17 & 36 & Belgium & 43,243 & 11.5 & 145 \\
7 & Iran & 5,252 & 82.0 & 40 & 37 & Brazil & 10,020 & 207.4 & 204 \\
8 & Israel & 39,974 & 8.3 & 97 & 38 & Canada & 44,773 & 35.6 & 307 \\
9 & Italy & 31,619 & 62.1 & 670 & 39 & Chile & 14,315 & 17.8 & 115 \\
10 & Kazakhstan & 8,585 & 18.6 & 10 & 40 & Czech Rep. & 19,818 & 10.7 & 125 \\
\midrule
11 & Korea, DPR & 1,360 & 25.2 & 18 & 41 & Denmark & 56,335 & 5.6 & 90 \\
12 & Korea Rep. & 29,730 & 51.2 & 158 & 42 & Egypt & 3,685 & 97.0 & 67 \\
13 & Mexico & 9,249 & 124.6 & 147 & 43 & Finland & 45,693 & 5.5 & 80 \\
14 & Mongolia & 3,553 & 3.1 & 6 & 44 & Norway & 73,615 & 5.3 & 41 \\
15 & Netherlands & 48,272 & 17.1 & 228 & 45 & Romania & 10,372 & 21.5 & 33 \\
16 & New Zealand & 41,629 & 4.5 & 35 & 46 & Serbia & 5,600 & 7.1 & 51 \\
17 & Poland & 13,429 & 38.5 & 162 & 47 & Slovak Rep. & 17,491 & 5.4 & 46 \\
18 & Portugal & 20,575 & 10.8 & 69 & 48 & Venezuela & 6,850 & 31.3 & 22 \\
19 & Russian Fed. & 10,248 & 142.3 & 436 & 49 & Uruguay & 17,252 & 3.4 & 5 \\
20 & Greece & 18,945 & 10.8 & 121 & 50 & Algeria & 4,225 & 41.0 & 2 \\
\midrule
21 & Hungary & 13,460 & 9.9 & 72 & 51 & Armenia & 3,690 & 3.0 & 28 \\
22 & South Africa & 6,089 & 54.8 & 122 & 52 & Azerbaijan & 4,098 & 10.0 & 10 \\
23 & Spain & 28,212 & 49.0 & 378 & 53 & Bulgaria & 7,924 & 7.1 & 67 \\
24 & Sweden & 53,248 & 10.0 & 145 & 54 & Colombia & 6,238 & 47.7 & 27 \\
25 & Switzerland & 80,837 & 8.2 & 138 & 55 & Estonia & 19,618 & 1.3 & 33 \\
26 & Thailand & 6,336 & 68.4 & 33 & 56 & Ireland & 68,604 & 5.0 & 50 \\
27 & Turkey & 10,434 & 80.8 & 80 & 57 & Malaysia & 9,660 & 31.4 & 10 \\
28 & Ukraine & 2,459 & 44.0 & 152 & 58 & Nigeria & 2,092 & 190.6 & 10 \\
29 & UK & 38,847 & 64.8 & 724 & 59 & Philippines & 3,022 & 104.3 & 5 \\
30 & Vietnam & 2,306 & 96.2 & 13 &  60 & Singapore & 53,880 & 5.9 & 2 \\
\bottomrule
\end{tabular}
\vspace{85mm} 
\end{table*}

\section{Results}
\subsection{GDP vs. IAU members}

\begin{figure*}
\centering
\includegraphics[width=150mm]{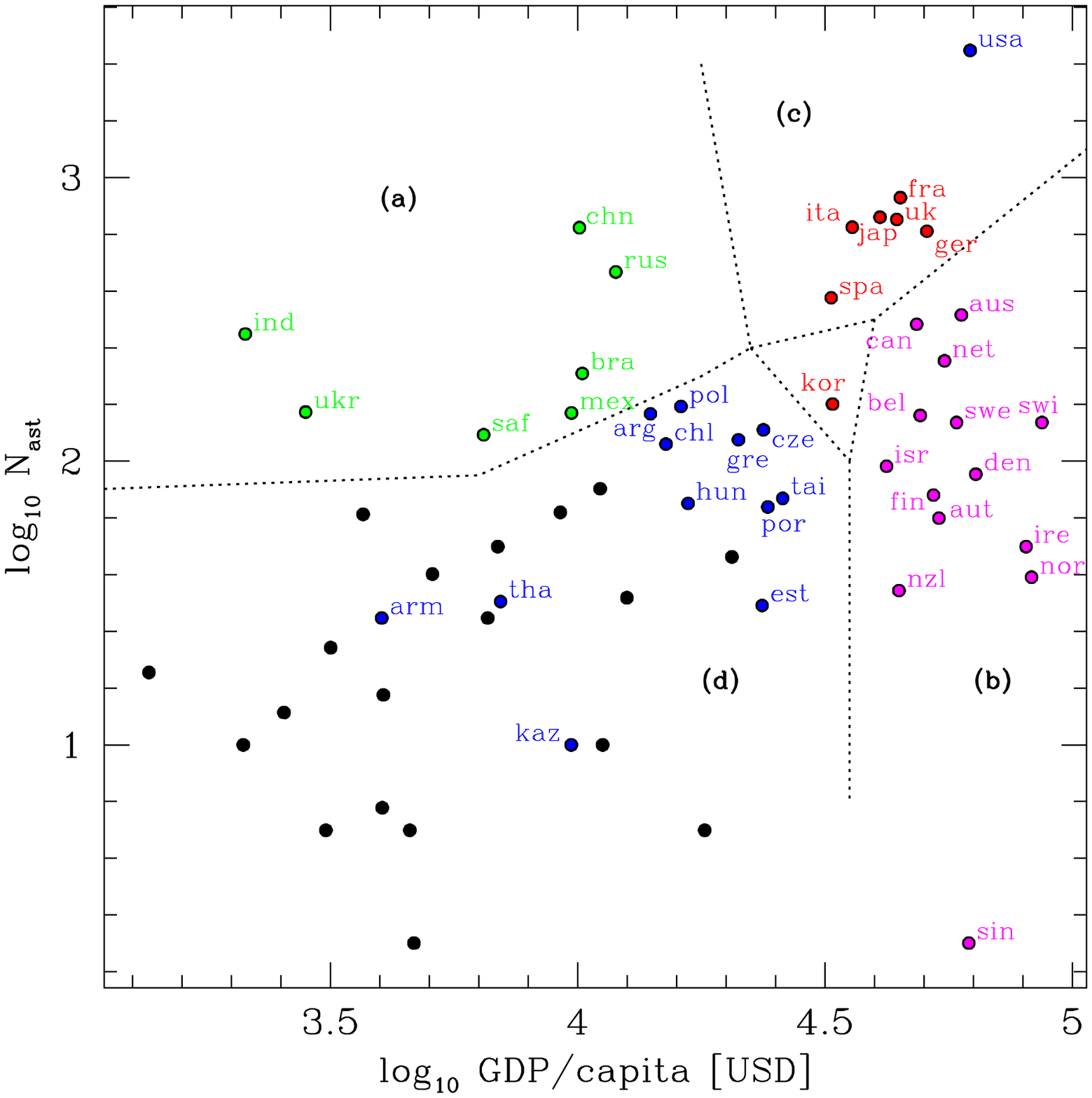}
\caption{{\bf GDP per capita versus number of IAU members.} The dotted lines are drawn to distinguish the BRICS countries denoted by green points in partition (a), the rich but relatively less populous countries denoted by magenta dots in partition (b), the developed and populous countries in Europe denoted by red dots in partition (c), and developing countries denoted by black dots in partition (d). The three alphabet letters represent the country names. Exceptionally, nkr represents the Democratic People’s Republic of Korea or North Korea, while kor represents the Republic of Korea or South Korea. Chn represents the People’s Republic of China, while chl represents Chile. Nzl stands for New Zealand.\label{fig:jkasfig1}}
\end{figure*}

At first, we plot the number of IAU members versus the GDP per capita in Figure~\ref{fig:jkasfig1} to see that there exist a number of groups: BRICS countries\footnote{BRICS is the acronym representing five major emerging economies: Brizil, Russia, India, China, and South Africa.}, rich but less populous countries, developing countries, and rich and populous countries. It is noteworthy that the Republic of Korea, denoted by kor, positioned at a crossroad in between the groups. 

\begin{figure*}
\centering
\includegraphics[width=150mm]{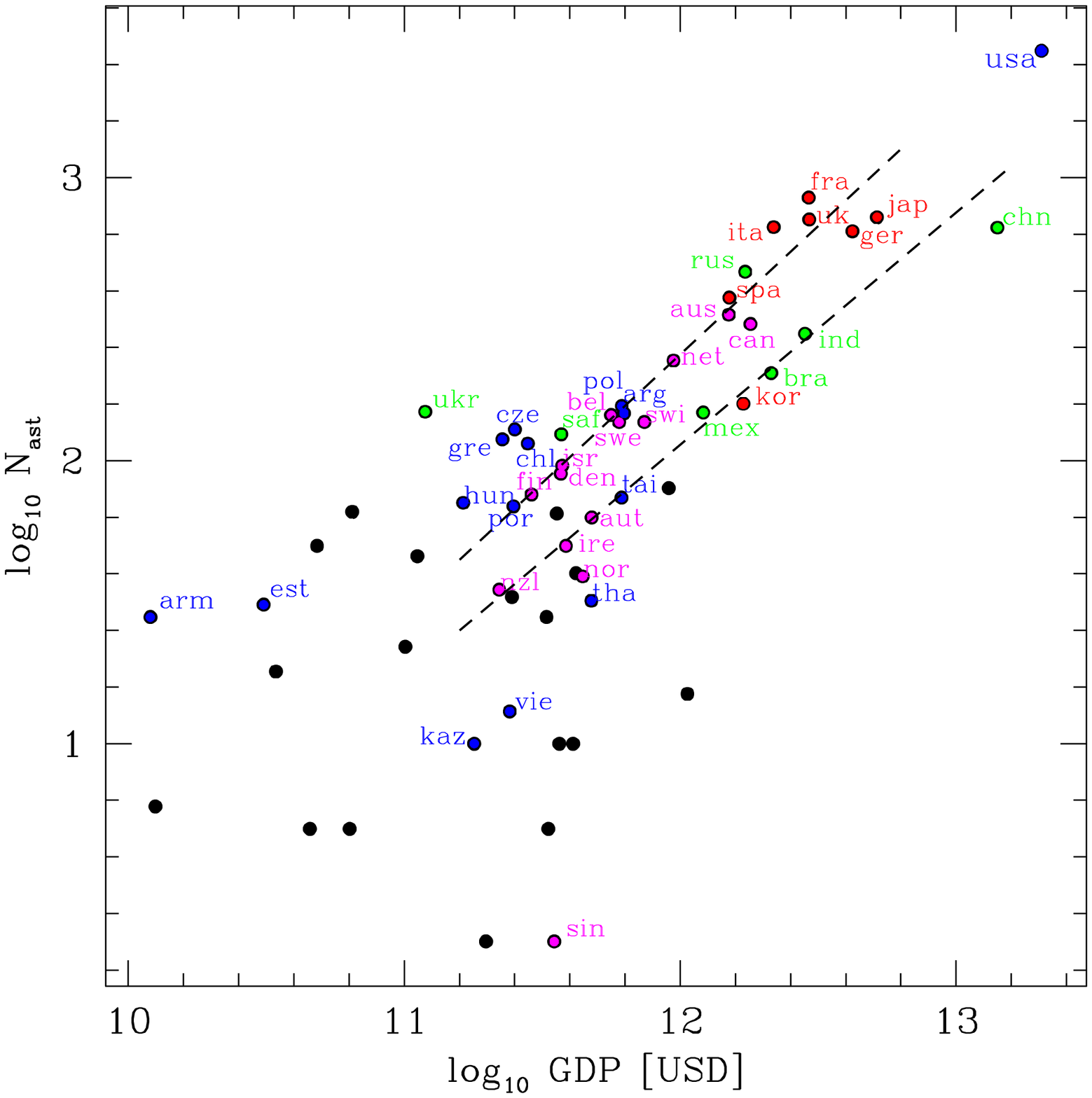}
\caption{{\bf GDP and number of IAU members} It shows a correlation between the economic capability and the size of astronomical community. The colours have the same meaning defined in Figure~\ref{fig:jkasfig1}. Two branches of correlations exist. The countries belonging to the upper branch have relatively long history of modern astronomy and developed economy, while those belonging to the lower branch have relatively short history of modern astronomy and relatively less-developed economy .\label{fig:jkasfig2}}
\end{figure*}

In order to see the correlation between economic power and size of astronomical community by country as a whole, the GDP and the number of IAU members are plotted in Figure~\ref{fig:jkasfig2}. Here, the color of each point agrees with those in Figure~\ref{fig:jkasfig1}. We confirm that the number of IAU astronomers in developed countries has a correlation with their GDP. Interestingly, the correlation in Figure~\ref{fig:jkasfig2} is branched into two groups, which are represented by the two dashed fitting lines. We find that \citet{kurtz05} performed the similar analysis for the data as of the year 2000 to find the two branches, which are separated by a factor of three in the GDP axis. We reach the same results, but our results are obtained by using the approximatedly-twenty-year-later data. Note that the axes of our Figure~\ref{fig:jkasfig2} is inverted comparing with their Figure 5. We make rough least square fits for the two branched data, and obtain the fitting line $y=0.91x-8.59$ for the upper line and $y=0.82x-7.78$ for the lower. Here $x\equiv \log_{10} GDP$ and $y\equiv \log_{10} N_{ast}$, where $N_{ast}$ is the number of astronomers. These correspond to a separation of two branches by a factor of 2.2 to 2.6 times in the GDP axis. We think that this factor is in rough agreement with the factor of three given by \citet{kurtz05}. However, it is not certain if there is any sophisticated change in the trends of the two branches during the last twenty years.

\Citet{kurtz05} also found that 31 European countries locate on the upper branch while only three on the lower branch, and that seventeen Asian countries locate on the lower branch. However, they did not take into account the South American countries such as Mexico and Brazil, which can be seen in Figure~\ref{fig:jkasfig2} of this paper and in their Figure 5. \citet{kurtz05} ascribed the polarization to the cultural difference between western nations and eastern nations, in terms of supports for basic science. Partially we agree on that opinion. However, we think that it would be rather proper to ascribe the the polarization to the difference of economic development stages of those countries.

\begin{figure*}
\centering
\includegraphics[width=150mm]{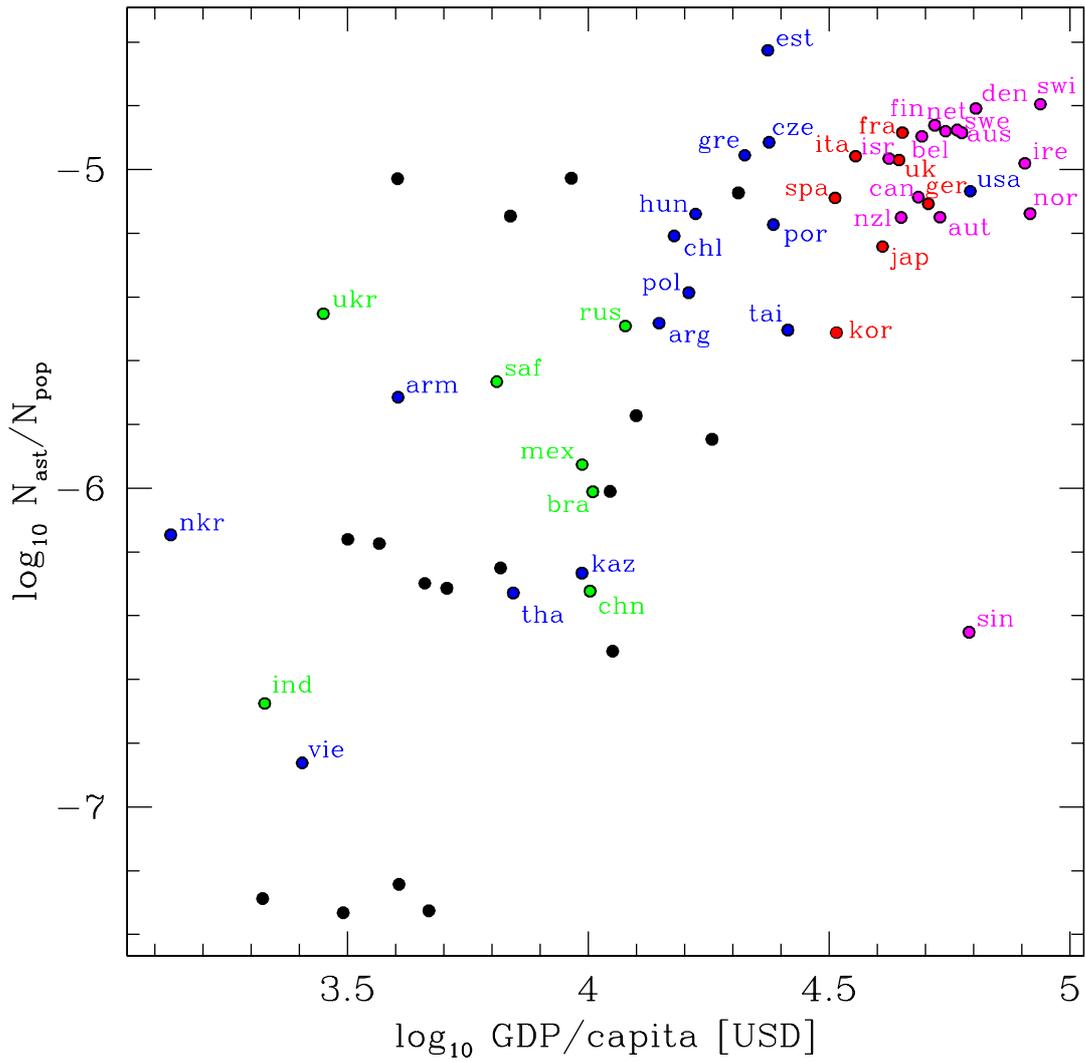}
\caption{{\bf GDP per capita and the number of IAU members per citizen.} The colours and the country names are the same to Figure 1. We can see the rough existence of two-branched correlations for the countries with relatively large GDP per capita.\label{fig:jkasfig3}}
\end{figure*}

The GDP per capita represents the average economic living standards of an individual citizen, and the number of IAU members per citizen indicates how much an individual citizen enjoys astronomy or is influenced by astronomical knowledge. We plot the GDP per capita and the number of IAU members per citizen in Figure~\ref{fig:jkasfig3}.  We can see that developed countries usually have approximately one astronomer per 100,000 citizens, while the so-called BRICS countries, marked with green dots, have approximately one astronomers per million citizens. We confirm that there is a rough correlation between GDP per capita and number of IAU members per one citizen. As was also pointed out in previous researches \citep{hearnshaw01,hearnshaw06}, there are a number of outliers. For example, Estonia is located far from the correlation in Figure~\ref{fig:jkasfig3}, meaning that the country has a relatively large number of astronomers compared to their GDP per capita. Another outliers such as Korea, Taiwan, Japan, Austria, and Norway have a relatively small number of astronomers compared to their GDP per capita values. In Figure~\ref{jkasfig3} we can roughly see the existence of two branches in the correlation. However, the constituent countries for the two branches do not coincide with those seen in Figure~\ref{fig:jkasfig2}. 

\subsection{Total Enumeration of Astronomers By Country\label{sec:global}}

As is mentioned above, the percentage of astronomers that have joined the IAU varies from country to country. Hence, in order to overcome this limitation, we have to count the total numbers of doctoral astronomers working in each country instead of just adopting the IAU members. We try to include astronomers, astrophysicists, and even physicists who study cosmology, high energy astrophysics, or astroparticle physics. However, engineers and technicians engaged in the development of instruments are not included in this paper, despite of their important contribution to astronomical researches.

Some countries like the United States\footnote{We also collect data from Astronomy and Astrophysics Survey Committee, {\it Status of Profession, in Working Papers: Astronomy and Astrophysics Panel Reports}, 321-325 (National Academy Press, 1991) \url{https://www.nap.edu/read/1635/chapter/1}; National Research Council, {\it in Federal Funding of Astronomical Research - chapter 4: Demographics.} 16-20, (National Academy Press, 2000); American Astronomical Society website: {\it Careers/Careers in Astronomy/Employment Potential}, \url{http://aas.org/learn/careers-astronomy/}} \citep{pold17}, 
the United Kingdom \citep{massey17,mcwhinnie17,murdin12}, 
Germany\footnote{Redaktionskomitee, {\it Deutsche Forschungsgemeinschaft Status und Perspektiven der Astronomie in Deutschland 2003-2016 Denkschrift} (ed. Burkert, A., Genzel, R., Hasinger, G., Morfill, G., Schneider, P. Koester, D.) 229-230 (WILEY-VCH Verlag GmbH \& Co. KGaA, 2003), \url{http://www.dfg.de/download/pdf/dfg_im_profil/geschaeftsstelle/publikationen/status_perspektiven_astronomie_2003_2016.pdf}; 
Redaktionskomitee des Rats deutscher Sternwarten, {\it Denkschrift 2017, perspektiven der Astrophsik in Deutschland 2017-2030: Von den Anfaengen des Kosmos bis zu Lebensspuren auf extrasolaren Planeten} (eds. Steinmetz, M., Brueggen, M., Burkert, A., Schinnerer, E., Stutzki, J., Tacconi, L. Wambsganss, J. \& Wilms, J.) 25-31 (Astronomische Gesellschaft, 2017), \url{http://www.denkschrift2017.de/Denkschrift2017.pdf}.}, 
Spain \citep{barcons07,gorgas16}, 
and Canada\footnote{Canadian Astronomical Society, {\it Unveiling the Cosmos: a Vision for Canadian Astronomy-Report of the Long Range Plan 2010 Panel}, p.13, p.83 (2011), \url{http://www.casca.ca/lrp2010/11093_AstronomyLRP_V16web.pdf}; Canadian Astronomical Society, 2015, {\it Unveiling the Cosmos: Canadian Astronomy 2016-2020, Report of the mid-Term Review 2015 Panel.} pp.104-106, \url{http://casca.ca/wp-content/uploads/2016/03/MTR2016nocover.pdf}; Racine, R. ``The evolution of astronomical and astrophysical populations in Canadian Universities'', a paper published on the following internet site, in which he provided statistical data of the results of the five astronomical and astrophysical censuses for the years of 1999, 2004, 2007-08, 2009-10, and 2013-14. \url{http://www.kcvs.ca/martin/astro/ecass/issues/2014-ve/features/racine/The\%20Evolution\%20of\%20A\&A\%20Populations.htm}} 
provided useful results of demographic surveys in either their long-term plans or annual reports, which can be used for estimating the numbers. 

We obtained the statistics for Gemany from Denkschrift 2003 and 2017 mentioned above, especially Table 2.2 of Denschrift 2017, published by the German astronomical society, but German academic ranking system is a bit special and so its equivalence is confirmed by a private communication \citep{steimetz18}. Japanese astronomical community provided the results of demographic surveys in the long-term plan\footnote{Astronomy and Space Physics subcommittee of Physics committee in Science Council of Japan, {\it Prospects and Long-term Planning of Astronomy and Astrophysics.} 14 (2010), \url{http://www.scj.go.jp/ja/member/iinkai/kiroku/3-0319.pdf}} \citep{sawa00}, and we also take into account the member statistics in the Bulletins of the Japanese Astronomical Society\footnote{Astronomical Society of Japan, {\it Annual Reports} (Astronomical Society of Japan, 2016), \url{http://www.asj.or.jp/asj/info/AnnReport2016.pdf}}. The number of astronomical scientists working in France as of 2003 was given by \citet{mamon03}, and the numbers as of 2018 are estimated by a private communication with him \citep{mamon18}.

The number of professors working in Korean universities is given in the long-term plan made by the Korean Astronomical Society \citep{lee17}. The number of astronomers in the institute, the Korea Astronomy and Space Science Institute, is obtained from the internal data\footnote{Human Resources Team of Korea Astronomy and Space Science Institute, 2018, a private communication}, the similar data for the universities is obtained by inspecting the Bulletins of the Korean Astronomical Society\footnote{The relevant annual reports have been published on the Bullentins of the Korean Astronomical Society 1990-2018. For example, see \citet{kas18}.}. 

The astronomical community of the Netherlands is composed of mainly four organizations \citep{boland13}: NOVA\footnote{In abbreviation of Nederlandse Onderzoekschool Voor Astronomie, NOVA web site: \url{http://nova-astronomy.nl/people/}}, ASTRON\footnote{In abbreviation of the Netherlands Institute for Radio Astronomy, ASTRON web site: \url{https://www.astron.nl/astronomy-group/people/people}}, SRON\footnote{Member list of each group in the SRON web site: \url{https://www.sron.nl/research}}, and JIVE\footnote{Joint Institute for VLBI ERIC (JIVE) web site: \url{http://www.jive.eu/staff}}. We counted only doctoral scientists directly from their web sites. In particular, the number of scientists in ASTRON at the end of 2017 was checked though a private communication\footnote{Steenbergen, A. 2018, A private communication. ``At the end of 2017, ASTRON employed 20 scientists (excluding R\&D engineers) on a permanent basis, and 28 scientists on fixed-term contracts. At the end of 2016, these figures were 14 and 31, respectively.''}, which gives a consistent result with our estimation. 

The Italian astronomical community has similar structure to the Korean one. The astronomers are working in either Instituto Nazionale di Astrofisica (INAF in abbreviation) or 25 universities, and the number of PhD scientists in the year of 2012 was reported in a paper \citep{sciortino13}. It is reasonable to assume that the number of astronomers in universities has not been changed much. We can find a detailed demographic information in the Astro-Dip database website\footnote{Astro-Dip Anagrafica dipendent, Database H1-HRMS \url{http://www.ced.inaf.it/anagrafica/}}, from which we can count the numbers of doctoral astronomers and astrophysicists working in the INAF. 

Similarly, the Taiwanese astronomical community provides a report of demographic investigation \citep{ip17}, we make a list of institutions and universities and count the number of PhD scientists one by one from their web sites. The number of doctoral astronomers in Australia was given in a paper written by \citet{sadler17}, and a relevant information can also be found in a decadal survey report published by the Australian Academy of Science in 2015\footnote{Australian Academy of Science, 2015, ``Australia in the Era of Global Astronomy: the Decadal Plan for Australian Astronomy 2016–2025'', Australian Academy of Science.\url{http://bit.ly/1P4Mfc5}}. All these numbers are faithfully cross-checked with the relevant literatures, but all procedures and more detailed descriptions are given in the Appendix section, and here we only show the data in Table~\ref{tab:jkastab2}. 

\begin{table*}
\caption{{\bf Total numbers of astronomers and astrophysicists, populations, GDP per capita, and GDP(PPP) per capita.} The populations and the GDP per capita are the same to Table 1, while the number of astronomers and astrophysicist are counted or estimated for the entire astronomical community.\label{tab:jkastab2}}
\centering
\begin{tabular}{lrrrrrrr}
\toprule
Country & Astronomers & Populations & GDP/capita & GDP/capita & PhDs/million  & IAU members & Ratio \\
 & (A) & (B) & USD & (PPP) USD & (A)/(B) & (C) & (C)/(A)\\
\midrule
USA	& 7,000 & 326.6 & 59,532 & 59,532 & 21 & 2,824 & 40\% \\
Germany & 1,400 & 80.6 & 44,470 & 50,639 & 18 & 654 & 47\% \\
Japan &1,500 & 126.5 & 43,279 & 43,279 & 12 & 728 & 49\% \\
UK & 1,400 & 64.8 & 40,412 & 42,656 & 26 & 724 & 52\% \\
France & 950 & 67.1 & 38,477 & 42,850 & 33 & 856 & 90\% \\
Italy & 1,000 & 62.1 & 31,953 & 39,427 & 16 & 670 & 67\% \\
Spain & 555 & 49 & 26,617 & 36,305 & 10	 & 378 & 68\% \\
Rep. of Korea & 310 & 51.2 & 29,743 & 38,335 & 6 & 158 & 51\% \\
Taiwan & 130 & 23.5 & 24,318 & 49,827 & 6 & 75 & 58\% \\
Netherlands & 390 & 17.1 & 45,638 & 52,503 & 23 & 228 & 58\% \\
Canada	 & 400 & 35.6 & 45,032 & 46,705 & 11 & 305 & 76\% \\
Australia &530 & 23.2 & 62,328 & 46,743 & 23 & 328 & 62\% \\
\bottomrule
\end{tabular}
\end{table*}

\begin{figure*}
\centering
\includegraphics[angle=0,width=150mm]{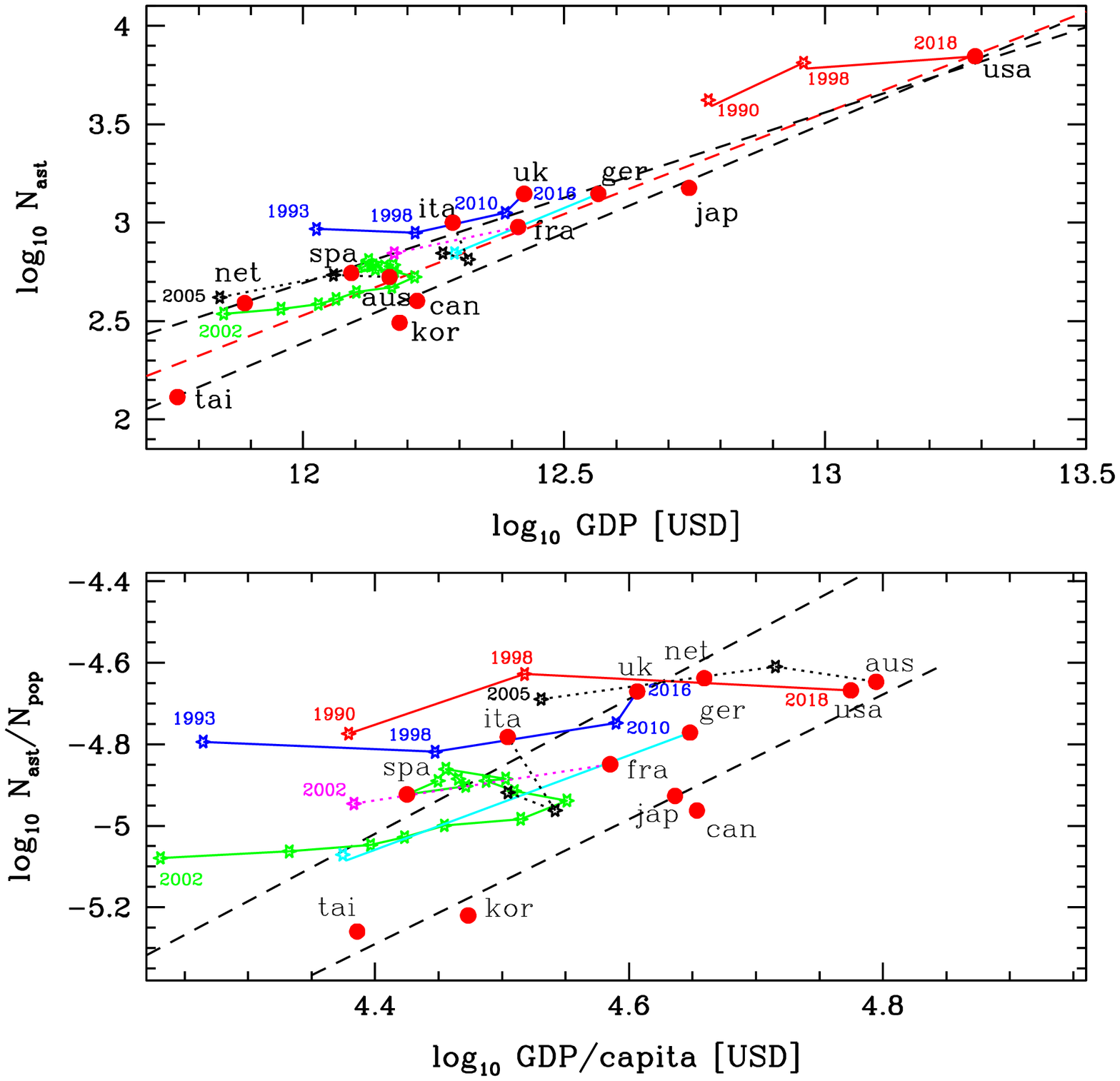}
\caption{{\bf GDP and the number of PhD astronomers.} The upper panel shows that the total numbers of astronomers with PhD degrees for a number of countries have a robust correlation with the GDP values. Note that the countries having relatively young history of modern astronomy show a correlation (lower black dashed line with a Pearson's correlation coefficient $R^2=0.98$) that is a bit different from the correlation seen in the European countries having relatively long history of modern astronomy (upper black dashed line with a Pearson's correlation coefficient $R^2=0.93$). The red dashed-line represents the linear regression for entire data with a Pearson's correlation coefficient $R^2=0.89$. Stars linked by lines represent the temporal variations of the relevant quantitites, which follow the correlations. The lower panel shows a rough correlation between the GDP per capita and the number of astronomers per citizen. The existence of two groups is also apparent. We see that most of advanced countries show rather constant number of astronomers, while the numbers of the Spanish, French, and the German astronomers have increased as their GDP values increased.\label{fig:jkasfig4}}
\vspace{25mm}
\end{figure*}

\begin{figure*}
\centering
\includegraphics[angle=0,width=150mm]{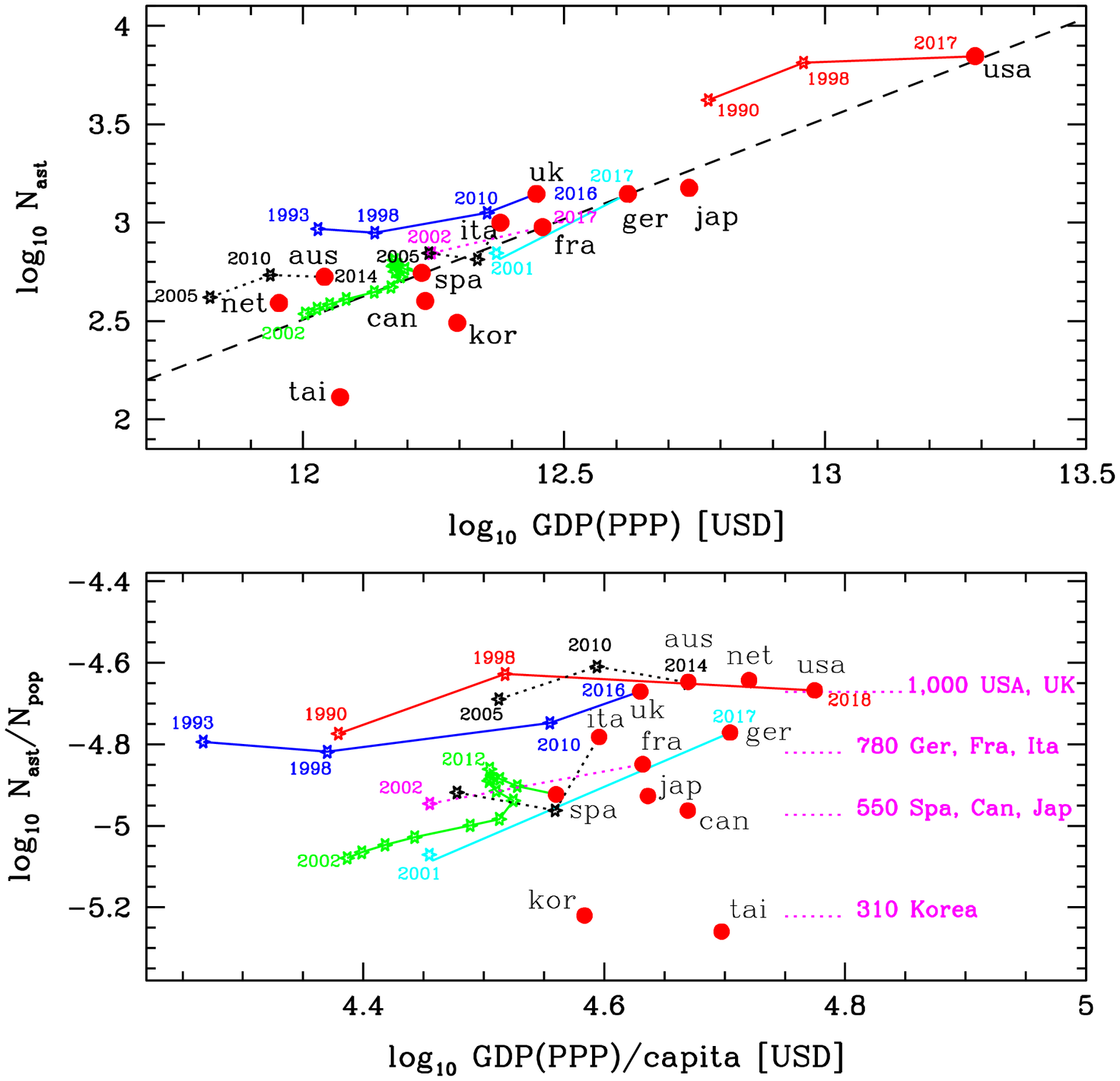}
\caption{{\bf GDP(PPP)/capita and the number of PhD astronomers} The same to Figure~\ref{fig:jkasfig4}, but we adopt the GDP based on Purchasing Power Parity per capita denoted by GDP(PPP)/capita in order to compare differences in living standards between nations. The GDP(PPP) and the number of astronomers show a rough correlation with a Pearson's correlation coefficient $R^2=0.75$, but the data excluding Republic of Korea (or South Korea), Taiwan, and Canada shows tighter correlation with a Pearson's correlation coefficient $R^2=0.93$. It is more apparent that the countries such as Republic of Korea, Taiwan, Cananda, Japan have relatively less astronomers and astrophysicists than expected considering their living standards.\label{fig:jkasfig5}}
\vspace{25mm}
\end{figure*}

In Table~\ref{tab:jkastab2}, we confirm our guess that the percentage of IAU membership varies from country to country. With the data in Table~\ref{tab:jkastab2}, we draw Figure~\ref{fig:jkasfig4} and Figure~\ref{fig:jkasfig5} in a similar manner to Figure~\ref{fig:jkasfig2} and Figure~\ref{fig:jkasfig3}. We can see the correlations similar to those in Figure~\ref{fig:jkasfig2} and Figure~\ref{fig:jkasfig3}. Interestingly, we can roughly confirm the existence of two groups in the correlations. Those countries including Australia, the USA, Japan, Canana, Korea, and Taiwan can be grouped as countries experienced recent rapid economic growth, while the others have long histories and have experienced recent gradual growth of the economy. 

In the previous research with the IAU membership, it was pointed out that Republic of Korea, Taiwan, Japan, Austria and Norway have a relatively small number of astronomers compared to their economic power \citep{hearnshaw06}. This fact cannot be ascribed to the low fraction of astronomers who had joined the IAU, because the fractions for those countries are not so much different from the other countries. We would rather pay attention to a fact that those countries such as Republic of Korea, Japan, Canada, Australia, and the USA have experienced relatively recent and rapid economic development in spite of their short history of modern astronomy, while the European countries having a relatively long history.

We also investigate temporal variation of the number of astronomers for several countries.
We see, in the upper panel of Figure~\ref{fig:jkasfig4}, a general trend that the number of astronomers has increased with the same rate of their GDP growths, which is conspicuous for countries such as the UK, Germany, France, Spain, the Netherlands, and the USA. What is more interesting is that the number of astronomers per citizen for the advanced countries, such as Australia, the USA, and the UK, has been nearly constant for the last 30 years, as can be seen in the lower panel of Figure~\ref{fig:jkasfig4}. Their GDP values and populations have surely been increased for that period. Hence, the total number of astronomers has also been increased as much.
On the other hand, France, Spain, Italy, and Germany show the relatively rapid growth of astronomers in the recent times.  In particular, Germany shows quite large increase in the number of astronomers per citizen, which might be caused by the governmentally-driven development of science after the German unification.

Until now we have adopted the GDP per capita in US dollars, which is just a nominal value. In order to compare the living standards one another in a more realistic manner, we adopt the gross domestic product based on purchasing power parity or GDP(PPP) in abbreviation. In Figure~\ref{fig:jkasfig5} we show the results obtained by adopting the GDP(PPP). Korea and Taiwan have relatively low living costs and so the GDP(PPP) values are larger than the nominal GDP values. Since the GDP(PPP) per capita values for the countries are larger, their living standards are nearly at the similar level to those of advanced countries. However, the numbers of astronomers, having PhD degrees and working in those countries, are relatively smaller than those of advanced countries. This means that those countries need more investment on astronomy. 

Based on the current number of astronomers per citizen of other advanced countries, we can estimate the proper number of astronomers working in Korea. In order for the Korean astronomical community to have comparable and competitive size to those of Spain, Canada, and Japan, the total number of astronomers working in Korea should be 550 as of 2018. 
If approximately 800 PhDs were working in Korea as of 2018, the Korean astronomical community could be comparable to those of Germany, France, and Italy. In order for the Korean astronomical community to be comparable to those of the USA, the UK, the Netherlands, and Australia, the total number of astronomers with PhD degrees and working in Korea should be approximately 1,000.
However, since the current number of astronomers, with PhD degrees and working in Korea, is at most 310, the Korean astronomical community should be able to create a large number of additional jobs in the near future as soon as possible in order to carry out fundamental and cutting-edge research projects and lead creative and meaningful discoveries.

\subsection{Korean Astronomical Community\label{sec:korea}}

We get to know that the current size of the Korean astronomical community is by far smaller than those of advanced countries. Then, we need to check the history of the Korean astronomical community and the current status to define a problem and to look for solutions by estimating the adequate number of astronomers working in the country.
 
\begin{table*}
\caption{{\bf Temporal increase of astronmers and astrophysicists working in Republic of Korea.} Data are gathered from Bulletins of the Korean Astronomical Society. Permanent positions and temporal positions are shown, and PDF means postdoctoral fellow. \label{tab:jkastab3}}
\centering
\begin{tabular}{lrrrrrrrr}
\toprule
year & \multicolumn{2}{c}{Universities} & \multicolumn{3}{c}{Institutes (mostly KASI)} & \multicolumn{3}{c}{Sum}\\
       & perm. & temp. & perm. & contract & PDF & perm. & temp. & total \\
\midrule
1990 & 24 & 7 & 8 & 0 & 0 & 32 & 7 & 39 \\
1992 & 25 & 0 & 10 & 0 & 0 & 35 & 0 & 35 \\
1994 & 42 & 3 & 16 & 0 & 0 & 58 & 3 & 61 \\
1996 & 43 & 4 & 20 & 1 & 0 & 63 & 5 & 68 \\
1999 & 50 & 14 & 25 & 4 & 0 & 75 & 18 & 93 \\
2000 & 46 & 20 & 27 & 5 & 0 & 73 & 25 & 98 \\
\midrule
2001 & 51 & 21 & 33 & 7 & 1 & 84 & 29 & 113 \\
2002 & 49 & 17 & 38 & 6 & 0 & 87 & 23 & 110 \\
2003 & 59 & 23 & 44 & 17 & 4 & 103 & 44 & 147 \\
2004 & 59 & 39 & 52 & 13 & 2 & 111 & 54 & 165 \\
2005 & 60 & 33 & 59 & 9 & 7 & 119 & 49 & 168 \\
2006 & 64 & 28 & 64 & 10 & 5 & 128 & 43 & 171 \\
2007 & 65 & 20 & 77 & 9 & 14 & 142 & 43 & 185 \\
2008 & 67 & 30 & 78 & 7 & 11 & 145 & 48 & 193 \\
2009 & 70 & 22 & 81 & 17 & 19 & 151 & 58 & 209 \\
2010 & 78 & 19 & 85 & 29 & 22 & 163 & 70 & 233 \\
\midrule
2011 & 77 & 29 & 91 & 39 & 26 & 168 & 94 & 262 \\
2012 & 79 & 37 & 97 & 35 & 14 & 176 & 86 & 262 \\
2013 & 77 & 46 & 105 & 56 & 20 & 182 & 122 & 304 \\
2014 & 75 & 57 & 117 & 55 & 14 & 192 & 126 & 318 \\
2015 & 77 & 51 & 127 & 56 & 8 & 204 & 115 & 319 \\
2016 & 77 & 44 & 129 & 56 & 10 & 206 & 110 & 316 \\
2017 & 82 & 42 & 138 & 38 & 6 & 220 & 86 & 306 \\
2018 & 86 & 50 & 147 & 20 & 7 & 233 & 77 & 310 \\
\bottomrule
\end{tabular}
\end{table*}

\begin{figure}
\centering
\includegraphics[angle=0,width=75mm]{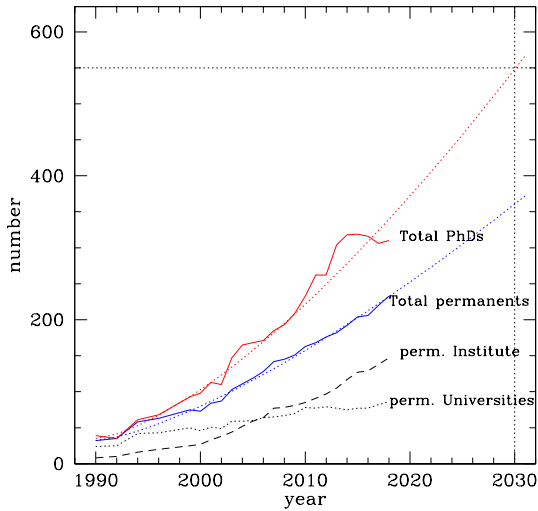}
\caption{{\bf Temporal variation of the number of astronomers in Korea for the last 30 years.} It is noteworthy that the number of astronomers working in the research institute has increased faster than that of professors in universities. The total permanent positions means the sum of the two categories. Adding the postdocs and contract positions to them makes the number of total PhD astronomers. If this trend continues, the number of PhD astronomers working in Korea will reach 550 that corresponds to the current size of astronomical communities for Spain, Canada, and Japan by the year of 2030.\label{fig:jkasfig6}}
\end{figure}

We collect the demographic data mainly from the previous issues of the Bulletin of Korean Astronomical Society and complemented by the Bulletin of Korean Space Sciences Society. Each member institution has reported their annual members and activities, and so we can trace the demographics back to early 1990s. We grouped the PhDs with permanent positions, either in Universities or in research institutes, and the others with non-permanent positions, either in contract positions or postdocs. 

We show the data in Table~\ref{tab:jkastab3}, and the temporal variations of number of PhDs in Figure~\ref{fig:jkasfig6}. We can see a rapid increase in the number of PhD astronomers in Korea for the last 30 years, which can be fitted with a function of $N(t)=2.35(t-1990)^{1.46}+35$ . The permanents in the research institute, a government-supported national institute called KASI has dominated the increase, and its workforce outnumbered professors in the Universities after the year of 2006 and becomes three times larger as of 2018. 
We can see that the number of PhDs working in Korea will be extrapolated to be approximately 550 by the year of 2030 if the inreasing trend will be continued. The number corresponds to the current sizes of astronomical communites such as Spain, Canada, and Japan.

\section{Conclusions\label{sec:con}}

In summary, we have found that the astronomical communities of countries around the world can be grouped into a few categories in the domain of the GDP per capita and the number of astronomers per citizen: i.e. large, populous, but less developed countries (or BRICS countries); highly matured, populous, and highly advanced countries; less populous but highly developed countries; developing countries. The Korean astronomical community lies inbetween them, which means that the Korean community is now at the crossroad of science development.

We have confirmed the correlation between the size of astronomical community and the economic power. Additionally, we confirm that the correlation for the countries having long tradition of modern astronomy is a bit different from that for the countries emerging later, which can be ascribed to a fact that some countries have experienced a recent economic development so rapidly  that their pure science sectors could not have been supported and developed in time. 

However, in general, a correlation may not imply causation. Hence, in order to check if any causation holds in this case, we assume that the larger economic capacity a country or the people has, the more supportive to astronomy they would be: that is, the number of astronomers working in the country must have been smaller in the past when the GDP of that country was lower. We obtain the temporal variation of the number of astronomers and the GDP values for a number of countries such as the United States of America, the United Kingdom, Germany, Australia, Spain, Italy, and France. From this statistics we have found that the number of astronomers is generally an increasing function of GDP or time. We also found that the number of astronomers for the countries having been experienced recent rapid economic growth, such as Germany, show rapid increases, while the countries having sufficiently developed economy, such as the United States, show relatively slow increase or nearly constancy in the number of astronomers per citizen. Thus, we conclude that the number of astronomers per citizen is a more important measure than other indicators to measure and estimate an adequate size of astronomical community.

Base on these observations, we have estimated a proper number of astronomers working in Korea considering GDP and population. We conclude that the Korean astronomical community could be competitive and comparable to those of Spain, Japan, and Canada if it had approximately 550 astronomers with PhD; it could be comparable to the German, French, and Italian astronomical communities if it had approximately 780 astronomers with PhD degrees; it could be comparable to the astronomical communities of the USA and the UK, if it had approximately 1,000 astronomers with PhD degree.

Subsequently, we have investigated the temporal variation of the size of the Korean astronomical community for the last 30 years. Although the community has experienced a rapid growth for the last three decades, the number of astronomers working in Korea is by far smaller than those of astronomers working in advanced countries when considering population and economic capability. The Korean astronomical community will be able to reach the capability of comminities such as Spain, Cananda, and Japan, if it grows in the similar trend  for the next decade.

A few problems related with our results will be discussed here, which must be solved in order to achieve the high level of astronomical sciences in Korea.
One problem is that more astronomers of the baby-boomer generation will retire during the next decade than the last decade. However, there are merely a few institutions that can produce PhDs. The current rate of PhD prodcution in Korea is at most 10 PhDs per year \citep{lee17}, including both domestic and abroad. However, we need at least double to triple rates in the next decade, when considering retirement rate. In spite of the circumstances, there are only 7 Universities having departments of astronomy and each of them is small in size. 
Another problem is the Korean education system. The number of professors that deliver astronomical knowledge to teachers has not grown as much as those teaching professional astronomers, which is also shown in Figure 5.4 of page 68 of \citet{lee17}. In other words, the number of professors in the University of education has not increased to a proper level. In fact, the number is nearly flat in recent years. Thus, astronomical knowledge cannot delivered to young students in elementary to high schools.
Moreover, astronomy classes are not widely taught in the majority of universities except for a few Universities having astronomy-related departments. Thus, the public understanding on astronomical knowledge is not so high in Korea. Surely, this stuation will eventually affect the procedures to decide the science policy, because the Republic of Korea is a democratic country. 

Another problem is that there is few occasions for the middle and high school students to do experiments in person during their school days. The students consume too much time to prepare the entrance examination for universities. The same is true among the undergraduates, to say nothing of the decreasing number of undergraduate studying science and engineering these days. The case in research institutes is not much different from these trends. The experimental parts of the institutes are relatively not strong, either. Although in this study we have only considered the numbers of astronomers and astrophysicists with PhD degrees, but it is noteworthy that there should be engineers and technicians as much as scientists that are carrying out astronomical researches to form a healthy ecosystem of scientific researches.

We will now consider the R\&D policy of the Korean government. It is well known that Repulic of Korea (or South Korea) shares a large amount of the budget to support the science and technology sector. According to a KISTEP\footnote{Korea Institute of S\&T Evaluation and Planning} statistics brief and the press release by the Ministry of Science, Technology, and ICT in July 2018 \citep{lee18}, the total amount of Korea’s R\&D investment reached 78.8 trillion won (69.73 billion USD) in 2017, up more than 13.5\% from that in 2016. The South Korean R\&D expenditure rankes fifth among OECD member countries after the United States, China, Japan and Germany. What is more interesting is that South Korea’s ratio of expenditure on R\&D to the GDP was 4.55\% in 2017, which is top higher than the rank 2 or Israel’s 4.25\%. A similar statistics as of 2015 can also be found in the Table 4-5 in the the Science and Engineering Indicators 2018 published by \citet{nsb18}, which means there has been a continuous support for the Korean science and technology sector by both the government and the private companies in recent years.

Then, what makes the pure-science sector represented by astronomy so vulnerable in Korea? We see, in the survey presented by the Korean government mentioned above, that the R\&D sources, either from the government or from the private companies, are not much different from other countries like the USA, Japan, Germany, France, the UK, and China. 
In many ways, it seems that the Korean science policy has been benchmarking that of Japan, as well as the USA.

We can doubt if the Korean astronomers spend less money than astronomers in other advanced countries. However, this is not the case. The R\&D expenditure per FTE (Full-time equivalent) in Korea is 182 thousand USD per year. (See Figure 3 in the page 4 of \citet{lee18}.) As of 2018, the Korean astronomical community has approximately 240 astronomers with permanent positions, as we have shown in Table~\ref{tab:jkastab3} of this paper, and they spend roughly 45 million USD per year. Hence, the R\&D expenditure per FTE per year is approximately 190 thousand USD per year. If we count all 310 PhDs, the R\&D expenditure per FTE per year is approximately 150 thousand USD per year, which is a bit smaller but nearly as large as to that of the UK.

There are two major differences in the Korean gross domestic R\&D expenditure structures from those of the advanced countries. First of all, the relative percentage of pure/basic science sector has been decreasing, especially in recent 5 to 10 years. (See Figure 11 on the page 8 of \citet{lee18}.) It is lower than the values of advanced countries such as France, the USA, and the UK. (See Figure 12 on the page 9 of \citet{lee18}.) Second of all, the expenditure for R\&D personnel is less than other advanced countries such as Germany and France, and larger than Japan and China. (See the Figure in the page 10 of \citet{lee18}.) Maybe this happens because the labor costs are different from country to country. In other words, the labor cost in Korea is cheaper than those in advanced countries. This also means Korean government did not hire as many astronomers as they can hire with the same amount of budget to the advanced countries. We have shown that the number of astronomers per citizen in Korea is too small to carry out outstanding researches. The most urgent thing is to increase the number of scientists in astronomy sector to meet the size of advanced countries. We needs vision to make the country strong in basic science.

Perhaps one of the most persuasive justifications for the investments to astronomy lie in the important role of astronomy that make the society adhering to innovative and cutting-edge sectors of science and technology by stimulating the deep desire of human beings to understand the origin, existence, and fate of the cosmos and human beings. We can find in history a large amount of examples that society’s support for merely curious-driven scientific researches has led to contributions to technology advances that has become long-term benefits to society. It is even said that the national investment to science is viewed as an essential element of economic strength and competitiveness. These concerns and interests all make people maintain the appropriately large number of astronomers in their countries, and in order to realize such consensus, a sufficient financial capacity is needed. Thus, we can see the correlation between the GDP and the number of astronomers, which can be a guide-line  for the emerging and developing countries.


\acknowledgments

The author is grateful to Prof. Hyung Mok Lee for useful discussions. He also thanks to the anonymous referee for the constructive comments that make the paper improved much.
This research was supported by Basic Science Research Program through the National Research Foundation of Korea(NRF) funded by the Ministry of  Science, ICT \& Future Planning(NRF-2018R1D1A1B07050035).
        

\appendix

\section{Detailed description on Census by Country\label{sec:data}}

{\bf USA} The 2016 demographic survey of the American Astronomical Society (hereafter abbreviated as AAS) showed that the entire AAS members consisted of 61\% full membership, 21\% junior members, 9\% associate members, 9\% emeritus, and 1\% educational affiliates \citep{pold17}. According to the same survey for the members whose addresses were registered in the United States, the percentage of respondents with doctoral degrees was approximately 80\% \citep{pold17}. As of the year 2018, the number of AAS members is about 7,000\footnote{American Astronomical Society website: Careers/ Careers in Astronomy/Employment Potential, \url{http://aas.org/learn/careers-astronomy/}}. Therefore, it is estimated that approximately 5,600 members of the AAS have PhD degrees. Moreover, there are about 1,000 astrophysicists who have exclusively joined the astrophysics division of the American Physical Society (hereafter APS in abbreviation), and there are also several hundred people who have not joined both societies. Thus, we estimate that there are approximately 7,000 doctoral astronomers or astrophysicists in the United States as of May 2018.

According to a survey by the National Research Council of the United States in 1998\footnote{Committee on Astronomy and Astrophysics, Board on Physics and Astronomy, Space Studies Board, National Research Council, 2000, {\it Federal Funding of Astronomical Research}, pp.16-20, Washington D.C.:National Academy Press. \url{http://nap.edu/9954}}, the AAS had 6,700 members, as of 1998. When extrapolating the fractional ratio of doctoral members, 80\%, in the year 2017 to the case of the year 1998, the number of doctoral members in AAS as of 1998 is estimated to be around 5,400. The same investigation shows that the APS had around 1,600 members in astrophysics in 1998. One third of them joined both societies, and so there were about 1,070 additional researchers joining exclusively in the APS. Therefore, by combining both societies, it can be estimated that in 1998 there were about 6,500 astronomers and astrophysicists working in the United States.
The Field Committee report in 1991 delivers that the United States had a pool of nearly 4,200 astronomers in 1990, up by 42 percent since 1980 as seen in the caption of Figure 1 in the report\footnote{Astronomy and Astrophysics Survey Committee, 1991, {\it Status of Profession, in Working Papers: Astronomy and Astrophysics Panel Reports}, pp.321-325, Washington D.C.:National Academy Press. \url{https://www.nap.edu/read/1635/chapter/1}}. Hence, there were about 3,000 astronomers around 1980.\\

{\bf Japan} In order to estimate the number of astronomers in Japan, we look up the membership of the Astronomical Society of Japan\footnote{Guide to Join, in the web site of the Astronomical Society of Japan, \url{http://www.asj.or.jp/asj/guide.html}} (hereafter abbreviated as ASJ). The members of the ASJ are divided into full members, associate members, group members, and supporting members. Among full members, there are student or graduate members who study astronomy or related disciplines. According to the 2016 Annual Report of the ASJ\footnote{Astronomical Society of Japan, Annual Report 2016, \url{http://www.asj.or.jp/asj/info/AnnReport2016.pdf}}, as of March 31, 2017, there were 2,059 full members including 506 student members, and 1,105 associate members. The full membership is defined as an individual who is engaged in astronomy and related fields and is responsible for the operation of the society\footnote{Regulations, in the web site of Astronomical Society of Japan, \url{http://www.asj.or.jp/asj/articles.html}}. In official documents that describes the long-term plan of the Japanese astronomical community, the number of full members is regarded as the number of approximate Japanese astronomers\footnote{Astronomy and Space Physics subcommittee of Physics committee in Science Council of Japan, 2010, {\it Prospects and Long-term Planning of Astronomy and Astrophysics}, p.14. \url{http://www.scj.go.jp/ja/member/iinkai/kiroku/3-0319.pdf}}. Therefore, the number of full members of ASJ, excluding the student members, is regarded as the number of PhD astronomers in Japan. The number for the year 2017 is approximately 1,500 persons.

We can obtain this number in another way. According to a survey of members of the ASJ published in January 2000 \citep{sawa00}, 607 persons among 1,316 respondents were doctoral researchers. Assuming that this ratio did not change much afterwards, we obtain the current number of doctoral astronomers in Japan to be 1,500, estimated from the fact that there are 3,243 members joined the ASJ as of March 31, 2017\footnote{Astronomical Society of Japan, Annual Report 2016, \url{http://www.asj.or.jp/asj/info/AnnReport2016.pdf}}.\\

{\bf United Kingdom} The Royal Astronomical Society (hereafter abbreviated as RAS) has published the demographic characteristics of RAS members as of autumn 2016 \citep{mcwhinnie17}. The Table 2 of that report provides us with the number statistics of staffs in astronomy and solar system sciences in 1993, 1998, 2010, and 2016. Thus, we can count the number of staffs working in universities and research establishments with interests in astronomy, solar system sciences, and cross disciplinary, excluding technical/support staffs and visitors. From the Table 1 of the report, we estimate the number-ratio of the staffs working in cross disciplinary fields to be 10\%. Thus, we obtain 1,000 scientists in 1993, 1,000 scientists in 1998, 1,200 scientists in 2010, and 1,400 scientists in 2016.\\

{\bf France} According to \citet{mamon03}, there were approximately 750 French scientists in the fields of astronomy and astrophysics, as of 2002, 44\% of them were working in the Centre National de la Recherche Scientifique (CNRS), 30\% were in the Observatories, 19\% were working in the Universities, and 7\% were hired by other institutions. \citet{mamon18} estimates approximately 800 full-time astronomers and astrophysicists in France, by extrapolating the net growth rate of tenure-track scientists and faculty: 15 posts per year from 1980 to 1987, 25 posts per year from 1988 to 2010, and 15 per year since 2011. In addition, by extrapolating the postdoc ratio of Institut Astrophysique de Paris (IAP) to other institutions of France, the number of postdocs working in France is estimate to be 135 postdocs. Hence, the number of doctoral scientists excluding engineers with PhDs to be approximately 935. Therefore, including retired and others it is estimated that there are approximately 1,000 astronomers or astrophysicists working in France, as of the year 2018.\\

{\bf Germany} A national member of the IAU on behalf of the German scientific community is the Rat Deutscher Sternwarten (hereafter abbreviated as RDS). According to Table 2.2 of Denkschrift 2017 or the 2017-2030 German Physics Development Plan published by RDS\footnote{Redaktionskomitee des Rats deutscher Sternwarten, 2017, {\it Denkschrift 2017, perspektiven der Astrophsik in Deutschland 2017-2030: Von den Anfaengen des Kosmos bis zu Lebensspuren auf extrasolaren Planeten}, eds. Steinmetz, M., Brueggen, M., Burkert, A., Schinnerer, E., Stutzki, J., Tacconi, L. Wambsganss, J. \& Wilms, J., pp.25-31, Potsdam:Astronomische Gesellschaft. \url{http://www.denkschrift2017.de/Denkschrift_2017.pdf}}, there are approximately 2,674 persons working in astronomy in Germany including 556 technical persons and 706 graduates (Promotion). Considering a German academic ranking system\footnote{Steinmetz, M. 2018, a private communication. ``The categories are roughly: W3/C4 represents full professors and directors, W2/C3 represents associate professors, AT/W1 represents assistant professors, E15 leaders of independent research groups, E13/E14/A13/A14 are scientists usually with a PhD degree, and Promotion are graduate students.''}, we estimate that 1,412 doctoral astronomers and astrophysicists were working in the German scientific community as of August 2017. Such demographic survey was also published in 2003 titled Deutsche Forschungsgemeinschaft Status und Perspektiven der Astronomie in Deutschland 2003-2016\footnote{Redaktionskomitee, 2003, Deutsche Forschungsgemeinschaft Status und Perspektiven der Astronomie in Deutschland 2003-2016 Denkschrift, ed. Burkert, A., Genzel, R., Hasinger, G., Morfill, G., Schneider, P. Koester, D., pp.229-230, Bonn:WILEY-VCH Verlag GmbH \& Co. KGaA. \url{http://www.dfg.de/download/pdf/dfg_im_profil/geschaeftsstelle/publikationen/status_perspektiven_astronomie_2003_2016.pdf}}. Its Tabelle 6.2 provides us the numbers of 674 scientific staffs in German institutes as of 2001. Its Tabelle 6.3 also gives us the numbers of even past times or 1962, 1987, and 1999. However, Germany was unified on 3 Oct 1990, and so it is not certain whether the numbers includes the east Germans or not. Hence, we adopt only the number 577 for the year of 1999, which is a sum of 375 scientists in Planstellen and 202 scientists in Drittmittel. It is remarkable that the number of astronomers and astrophysicists in Germany has been increased rapidly during the 21st century.\\

{\bf Spain} According to a recent demographic survey for 50 research institutes in Spain, as of 2016, there were 391 professional astronomers, 210 postdocs, and 186 graduates in Spain\citep{gorgas16}. In the presentation slide for Segundo informe de los recursos humanos en Astronomia y Astrofisica en Espana\footnote{Sociedad Espanola de Astronomia, Evolucion del personal in presentation slides titled Segundo informe de los recurso humanos en Astronomia y Astrofisica en Espana, \url{https://www.sea-astronomia.es/sites/default/files/informe_2016_v1.pdf}}, the number statistics of human resources from 2002 to 2016 are given. We show the table in Table~\ref{tab:jkastab4} of this paper. We count the number of scientists classified as Plantilla (posts or staffs) and Postdocs and exclude the number of predoctors to obtain the number of PhDs. We obtain the number of doctoral astronomers or astrophysicists working in Spain to be 555 as of 2016.\\

\begin{table}
\caption{{\bf Temporal changes of human resources of Spanish astronomical community.} The data is obtained from the presentation slide for ``Segundo informe de los recursos humanos en Astronomia y Astrofisica en Espana''.\label{tab:jkastab4}}
\centering
\begin{tabular}{ccccc}
\toprule
 & Plantilla & Postdocs & Predocs & Total \\
 \midrule
2002 & 233 & 112 & 130 & 475 \\
2003 & 242 & 123 & 141 & 506 \\
2004 & 251 & 134 & 159 & 544 \\
2005 & 261 & 148 & 173 & 582 \\
2006 & 279 & 166 & 190 & 635 \\
2007 & 291 & 179 & 185 & 655 \\
2008 & 306 & 224 & 209 & 739 \\
2009 & 317 & 247 & 216 & 780 \\
2010 & 320 & 280 & 216 & 816 \\
2011 & 334 & 276 & 218 & 828 \\
2012 & 348 & 297 & 229 & 874 \\
2013 & 343 & 265 & 218 & 826 \\
2014 & 341 & 241 & 214 & 796 \\
2016 & 347 & 208 & 185 & 740 \\
\bottomrule
\end{tabular}
\end{table}

{\bf Italy} According to \citet{sciortino13}, scientists carrying out astronomical research in Italy were working either within the Instituto Nazionale di Astrofisica (abbreviated as INAF) or within research groups in about 25 universities, and there is a small number of INFN(Italian Institute for Nuclear Physics) scientists doing researches on astro-particle physics or related topics. He also said that these scientists were either permanent or temporary staff members, or junior, even senior postdocs. Thus, we can regard the number of these scientists as that of astronomers and astrophysicists with a PhD degree working in Italy. He reported that, as of June 2012, there were 450 staff scientists, 70 temporary staffs, and 270 postdoctoral researchers in INAF, and also that there were 95 professors, 73 researchers, 30 postdoctoral researchers, and approximately 90 graduates working in the universities \citep{sciortino13}. Adding them all, we estimated that there were 980 doctoral scientists working in astronomical community of Italy as of June 2012.
He also mentioned that there were 1,037 permanent staff employees in INAF as of 2005, and 540 among them were scientists or technologists, which is a slightly larger number than that of 2012. Assuming the numbers of both postdocs in INAF and doctoral scientists in universities are not changed a lot, we estimate the number of astronomers and astrophysicists working in Italy to be ~1,000 as of 2005.

According to the INAF website\footnote{Astro-Dip Anagrafica dipendent, Database H1-HRMS, \url{http://www.ced.inaf.it/anagrafica/}}, as of November 2018, the INAF has a total of 1,153 employees, and 745 persons among them are research personnel, and 408 persons are technical or administrative personnel. There are 466 permanent staff researchers, 61 temporary staff researchers, and 271 postdocs or fellows. If we can assume the human resources in the universities did not change much, we can estimate the number of their researchers to be ~190. Hence we estimate that ~1,000 doctoral astronomers are working in Italy, as of 2018.\\

{\bf the Republic of Korea} In the Republic of Korea or South Korea, approximately two third of doctoral astronomers are working in the Korea Astronomy and Space Science Research Institute (KASI in abbreviation), and the other one third are working in a few universities. As of November 2018, the Korea Astronomy and Space Science Institute has 147 full-time employees and 27 fixed-term researchers and postdocs with doctoral degree including approximately 10 administrative staffs with doctoral degrees, and 4 research fellows\footnote{Size of human resources, Management information system in the internal web site of the Korea Astronomy and Space Science Institute.}. Thus, KASI has approximately 168 doctoral researchers. According to Bulletins of Korean Astronomical Society, as of April 2018, 76 full-time professors, 15 research professors, and 35 postdoctoral researchers are studying in the universities \citep{kas18}. About ten additional PhDs are counted in other research institutes, science high-schools, science museum, and private companies. Hence, we see that there are approximately 310 doctorate astronomers working in Republic of Korea.\\

{\bf The Netherlands} According to a survey on the status of astronomy in the Netherlands \citep{boland13}, as of 2012, the Dutch astronomers belonged to the Nederlandse Onderzoekschool Voor Astronomie (abbreviated as NOVA), the Netherlands instituut voor radioastronomie (abbreviated as ASTRON) investigating radioastronomy and pure astronmy, Stichting Ruimte Onderzoek Nederland (abbreviated as SRON), and VLBI ERIC Joint Institute for VLBI ERIC (abbreviated as JIVE). Since these four organizations provide member lists in their website, we count the number of doctoral scientists in the employee lists. We also visit some of the personal websites to check whether he or she has a PhD degree. 

As of 2018, four universities participate the NOVA: Amsterdam, Groningen, Nijmegen, and Leiden\footnote{NOVA web site, \url{http://nova-astronomy.nl/people/}}. Amsterdam has 54 PhDs among 161 employees, 49 doctoral students and 45 master students. Groningen has 40 doctors among 93 employees and 11 PhD students. Nijmegen has 31 doctors among 68 employees, 21 doctoral students, and 8 master students. Leiden observatory has 116 PhDs among 450 employees, 89 doctoral students, and 90 master students. Thus, there are 241 doctors in NOVA as of 2018.

According to NOVA's 2015 annual report\footnote{NOVA Annual Report 2013-2014-2015, \url{http://nova-astronomy.nl/wp-content/uploads/2016/10/Annual-Report-2013-2014-2015-PDF.pdf}}, the NOVA members consisted of approximately 56 FTE senior staff members in permanent and tenure-track positions, approximately 10 FTE senior postdocs, 83 FTE postdoctoral fellows, approximately 40 instrumentalists, 171 PhD students, and 5 FTE co-workers from ASTRON and SRON. Thus, the number of astronomers with PhD degrees involved with the NOVA program was approximately 150 as of 2015. 

ASTRON has four groups including astronomy group, radio observatory, R\&D labs, and NOVA IR/Optical\footnote{ASTRON web site: \url{https://www.astron.nl/astronomy-group/people/people}}. Excluding technological staffs and administrative staffs, ASTRON has 52 PhDs among 146 employees. According to a private communication with Anneke Steenbergen\footnote{Steenbergen, A. a private communication. ``At the end of 2017, ASTRON employed 20 scientists (excluding R\&D engineers) on a permanent basis, and 28 scientists on fixed-term contracts. At the end of 2016, these figures were 14 and 31, respectively.''}, there were 48 scientists employed excluding R\&D engineers, 20 having permanent positions and 28 having fixed-term contracts as of 2018, which is not so different from the direct counting from the web site. SRON has six subgroups including Earth, Exoplanet, Astrophysics, Technology, Instrument, engineering groups\footnote{Members list of each group in the SRON web site: \url{https://www.sron.nl/research}}. It has 89 PhDs among 204 employees. JIVE has 10 doctors among 26 employees\footnote{Joint Institute for VLBI ERIC (JIVE) web site: \url{http://www.jive.eu/staff}}. Hence, the total number of PhDs working in these three organisations is estimated to be 151 as of 2018.

If all being combined, it is estimated that there are approximately 390 doctoral researchers investigating astronomy and astrophysics in the Netherlands. There are also highly educated engineers working in the Dutch astronomical community as much as this number. It is also remarkable that the Netherlands host a couple of huge centers for astronomical researches and space exploration in Europe.\\

{\bf Canada} According to a survey by the Association of Canadian Universities for Research in Astronomy (abbreviated as ACURA), as of the year 2010, the Canadian astronomical community consists 200 professor-level researchers and 100 postdocs, and 300 graduate students\footnote{Canadian Astronomical Society, 2011, ``Unveiling the Cosmos: a Vision for Canadian Astronomy - Report of the Long Range Plan 2010 Panel'', p.13, p.83. \url{http://www.casca.ca/lrp2010/11093_AstronomyLRP_V16web.pdf}}. According to an updated survey of ACURA, as of 2013, 17 professors, 37 postdocs, and 71 graduate students were increased\footnote{Canadian Astronomical Society, 2015, ``Unveiling the Cosmos: Canadian Astronomy 2016-2020'', Report of the mid-Term Review 2015 Panel, pp.104-106. \url{http://casca.ca/wp-content/uploads/2016/03/MTR2016nocover.pdf}}. Thus, there were approximately 350 PhDs as of 2013, and the growth rate was only 3\%. Adopting this growth rate, the number of doctoral astronomers is extrapolated to be approximately 400 as of 2018. 

On the other hand, according to Dr. Racine’s research\footnote{Racine, R. ``The evolution of astronomical and astrophysical populations in Canadian Universities'', \url{http://www.kcvs.ca/martin/astro/ecass/issues/2014-ve/features/racine/The\%20Evolution\%20of\%20A\&A\%20Populations.htm}}, the number of tenure stream faculty members, adjuncts and postdocs were 164 as of 1999, 247 as of 2004, 285 as of 2008, 305 as of 2010, 341 as of 2014. If we also apply the manpower growth rate of 3\% and extrapolate, then we can estimate the number of PhD astronomers and astrophysicists working in Canada to be 410 as of 2018, which is close to the numbers estimated by the demographic survey by ACURA. In conclusion, we estimate the number of PhD astronomers and astrophysicists working in Canada is approximately 400, as of 2008 Nov.\\

{\bf Australia} The number of doctoral astronomers in Australia is given in \citet{sadler17}. According to this paper, the number of astronomy jobs in Australia have been changed from 417 in 2005 to 542 in 2010 and to 527 in 2014. In 2014, the percentage of full-time astronomers is 64\%, so the number of doctoral researchers is estimated to be approximately 530, among which 340 are full-time astronomers.\\

{\bf Taiwan} In order to estimate the number of doctoral astronomers in Taiwan as of 2018, the professors, researchers, and postdoctoral researchers working in Taiwan's major astronomical institutes and universities, as mentioned in a paper on the history of Taiwanese astronomy \citep{ip17}, were surveyed in their websites. Institute of Astronomy in National Ting Hua University has 11 doctoral researchers, National Taiwan University has 4 doctoral researchers, National Taiwan Normal University has 8 doctoral researchers, National Central University has 17 doctoral researchers, Academia Sinica Institute of Astronomy and Astrophysics has 83 doctoral researchers, and Academia Sinica Institute of Physics has 4 doctoral researchers working in Taiwan. Thus, the number of doctoral researchers working in astronomy and astrophysics in Taiwan is estimated to be approximately 130.


\end{document}